\documentclass[referee]{aa}

\usepackage{txfonts}
\usepackage {graphicx}
\usepackage{placeins}
\usepackage {natbib}
\bibpunct{(}{)}{;}{a}{}{,}
\usepackage{longtable,array,tabularx}

\begin{document}
\title{Near infrared imaging of the broad absorption line quasar \\ \object{BAL QSO~0134+3253}
\thanks{Based on observations with the VLT; proposal number 67.B-0019}}
\author{J. Zuther, A. Eckart}
\institute{I. Physikalisches Institut, Universit\"at zu K\"oln,
Z\"ulpicher Str. 77, 50937 K\"oln}
\offprints{J. Zuther, \email{zuther@ph1.uni-koeln.de}}
\date{Received 2005 March 8 / Accepted 2005 August 5 }
\abstract{In this paper we present near infrared (NIR) imaging data of the host galaxy of the broad absorption line quasar (BALQ) at $z=2.169$, serendipitously found close to 3C~48 \citep{1998AJ....115..890C}. The data were obtained with the ESO-VLT camera ISAAC during period 67. We find extended, rest-frame optical emission around the BALQ after subtracting a scaled stellar point spread function from the quasar nucleus in $J$, $H$, and $Ks$. The extended rest-frame optical emission can be interpreted as an approximately 2~Gyr old stellar population composing the host galaxy of the BALQ or a stellar population of similar age associated with an intermediate ($z=1.667$) absorption system spectroscopically identified by \cite{1998AJ....115..890C} simultaneously. The rest-frame-UV emission on the other hand is dominated by a young, $\sim$ 500~Myr old stellar population \citep{1998AJ....115..890C}. The UV/optical colors resemble a mixture of the two populations, of which the young one accounts for about 80\%. Assuming that the residual emission is located at the BALQ redshift, we find that the host galaxy has a resolved flux of about 10\% of the BALQ flux. The physical scale is quite compact, typical for radio quiet QSOs or Lyman break galaxies at these redshifts, indicating that the systems are still in the process of forming.
}
\authorrunning{J. Zuther et al.}
\titlerunning{$JHKs$ imaging of BALQ~0134+3253}
\maketitle


\section{Introduction}
There is evidence for an intimate link between the growth of massive black holes and the galaxy bulges they reside in \citep[e.g.][]{2001Sci...294.2516P}. This implies the existence of physical feedback mechanisms to regulate their coeval evolution. In the quasar phase, i.e. during the bulk of the accretion of matter onto the central massive black hole, winds/outflows from the innermost region of the quasar can be an important source of such a feedback \citep[e.g.][]{2004AdSpR..34.2594G}.

Broad absorption line quasars (BALQs) form a rare class, comprising about
20\% of the QSO population at moderate to high redshifts \citep{2003AJ....125.1784H}. The BAL phenomenon is related to outflows in quasars \citep[e.g.][]{2004ASPC..311..203H}. Such outflows constitute metal absorption systems with large blueshifted velocities (several thousand km~s$^{-1}$). 
There are currently two standard interpretations of these phenomena. \emph{(1) Unified scheme}: Under the assumption of the unified scheme for active galaxies \citep{1993ARA&A..31..473A}, the absorbing  clouds have a small covering factor as seen from the QSO nucleus \citep{1991ApJ...373...23W}. Thus, the frequency of detection just translates to the rate at which our line of site intercepts the outflow.
\emph{(2) Evolutionary scheme}: A different interpretation is that these objects are very young and are ejecting their gaseous envelopes at very high velocities following the initial turn-on of the active galactic nucleus \citep{1984ApJ...282...33H,1984ApJ...287..549B}.

The unified scheme would imply that all radio-quiet QSOs should be classified as BALQs if observed from the proper angle. This was supported until recently by the work of \citep{1992ApJ...396..487S}. However, in a spectroscopic follow-up of radio selected BALQs, \cite{2000ApJ...538...72B} find a population of FIRST\footnote{VLA Faint Images of the Radio Sky at Twenty-centimeters} BALQs being rather radio-intermediate, indicating that the radio properties do not support the "simple" scenario stated above. 

\begin{figure*}[t!]
\resizebox{\hsize}{!}{\includegraphics{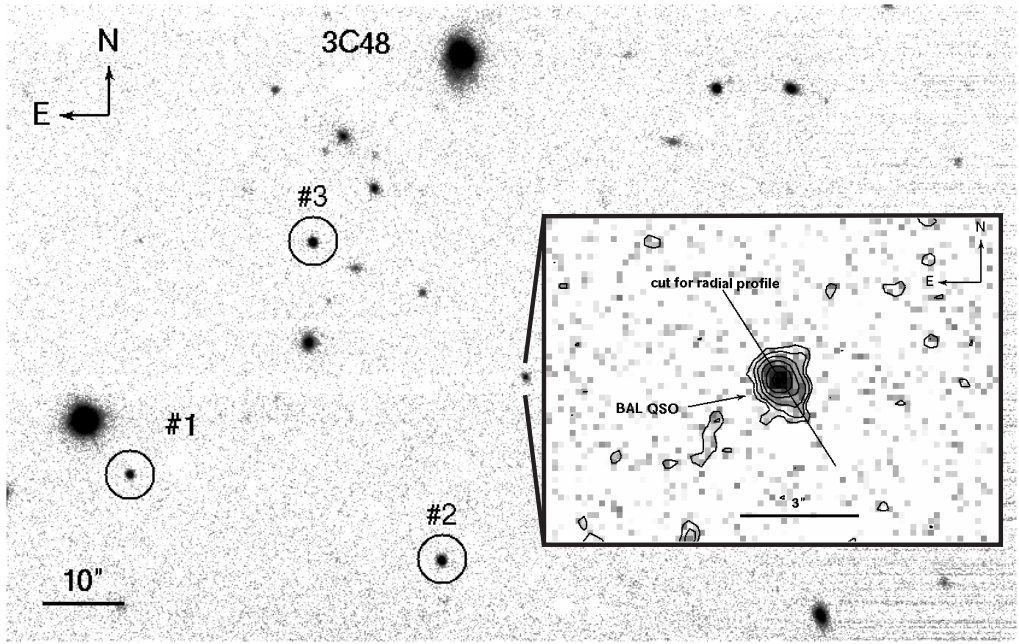}}
\caption{$H$ band ISAAC image of the field of 3C~48 \citep{2004A&A...414..919Z}.The inset shows our $H$ band close up of the broad absorption line quasar. The lowest contour is at the level of sky noise. Also indicated is the primary cut used for the radial profiles. The stars used for PSF estimation are marked with (\#1, \#2, and \#3). See text for details.}
\label{fig:hbal}
\end{figure*}
\defcitealias{1998AJ....115..890C}{CS98}

On the other hand, in case of the evolutionary scenario an enhanced sub-mm radiation would be expected, because the BALQs still possess a lot of molecular gas and dust while actively forming stars \citep{1997ApJ...484L..17H,2003AJ....126.2594R,2003AJ....126.1131R} and the host galaxies would appear to be at early stages in their evolution. There seems to be a subclass of BALQs with enhanced far-IR emission \citep{1989ApJ...340L...1L} for which this could be the case. The large number of low-ionization BALQs in IRAS-selected luminous QSOs, along with the reddening from dust in these objects, suggests that this subclass may represent an evolutionary transition between ultraluminous IR galaxies and the classical QSO population in the early universe \citep{1994ApJ...436..102L,1993ApJ...413...95V,1996AJ....112...73E}.
Recent studies of BALQs by \cite{2003ApJ...596L..35L} and \cite{2003ApJ...598..909W}, however, show no differences in the sub-mm properties between BALQs and non-BALQs which supports a small BAL covering factor.  

In addition \cite{2002MNRAS.336..353A} show that the star forming phase of the host and the dust/gas rich BAL wind do not need to coincide in time, thus providing constraints on alternative evolutionary models. 

As the detailed nature of these objects is still under debate, the study of their host galaxies could provide important complementary information.

One recent example of a detected extended emission around a BAL QSO is the object 0134+3253. \cite{1998AJ....115..890C} (hereafter CS98) present in their work a rest frame UV spectrum as well as Hubble Space Telescope (HST) rest frame UV/optical images of the $z=2.169$ BALQ~0134+3253. The authors find a deep absorption trough blueward of the \ion{C}{IV} line which is composed of three elements and wind velocities of up to -12,000~km~s$^{-1}$. The red edge is about +1000~km~s$^{-1}$ (their Fig. 2). There is no more spectroscopic information available to further classify this QSO as an high- or low-ionization BALQ. From their optical spectrum an intermediate absorption system at $z\approx 1.667$ is evident, probably being a damped Ly$\alpha$ absorber  \citepalias{1998AJ....115..890C}. In this context it is not clear whether the redness with respect to the average HST quasar spectrum \citep{1997ApJ...475..469Z} is due to the absorption system or intrinsic to the quasar host galaxy. Photometric data are summarized in Tab. \ref{tab:prevBalPhoto}. The HST F555W and F814W filters approximately correspond to $V$ and $I$ band, respectively, whereas $g$ lies between $B$ and $V$.

In section 2 we will briefly describe the ISAAC observations. In section 3 we will present rest frame optical photometry of this BALQ and its extended emission to complement the rest frame UV data of \citetalias{1998AJ....115..890C}. Unless noted, we use a cosmology with $H_0=70$~km~s$^{-1}$~Mpc$^{-1}$, $\Omega_m=0.3$, and $\Omega_\Lambda=0.7$ \citep{2003ApJS..148..175S} throughout this paper. 

\begin{table}[ht]
\caption{Previous measurements of the BALQ and the extended emission.}
\label{tab:prevBalPhoto}
\begin{center}                       
\begin{tabular}{ccc}
\hline
\hline
Filter      & Total & Host\\
(1) & (2) & (3)\\
\hline
$g$         & 21.2 $^\mathrm{a}$  &--\\
F555W $=V$  & 21.14 $^\mathrm{b}$ & $23.8\pm 0.1$ $^\mathrm{b}$\\
F814W  $=I$ & 21.28 $^\mathrm{b}$ & $24.1\pm 0.1$ $^\mathrm{b}$\\
$H+K$       & -- & 5 $\times 10^{-20}$~erg~s$^{-1}$cm$^{-2}$\AA$^{-1}$arcsec$^{-2}$ $^\mathrm{b, c}$\\
\hline
\end{tabular}
\end{center}
\begin{list}{}{}
\item[$^{\mathrm{a}}$] \cite{1994PASP..106..646K}; listed as a star
\item[$^{\mathrm{b}}$] \citetalias{1998AJ....115..890C}; measured in a 2\farcs 6 diameter aperture
\item[$^{\mathrm{c}}$] \citetalias{1998AJ....115..890C}; $1\sigma$ detection threshold 
\end{list}
\end{table}

\begin{figure*}[ht!]
\begin{center}
\resizebox{\hsize}{!}{
\includegraphics{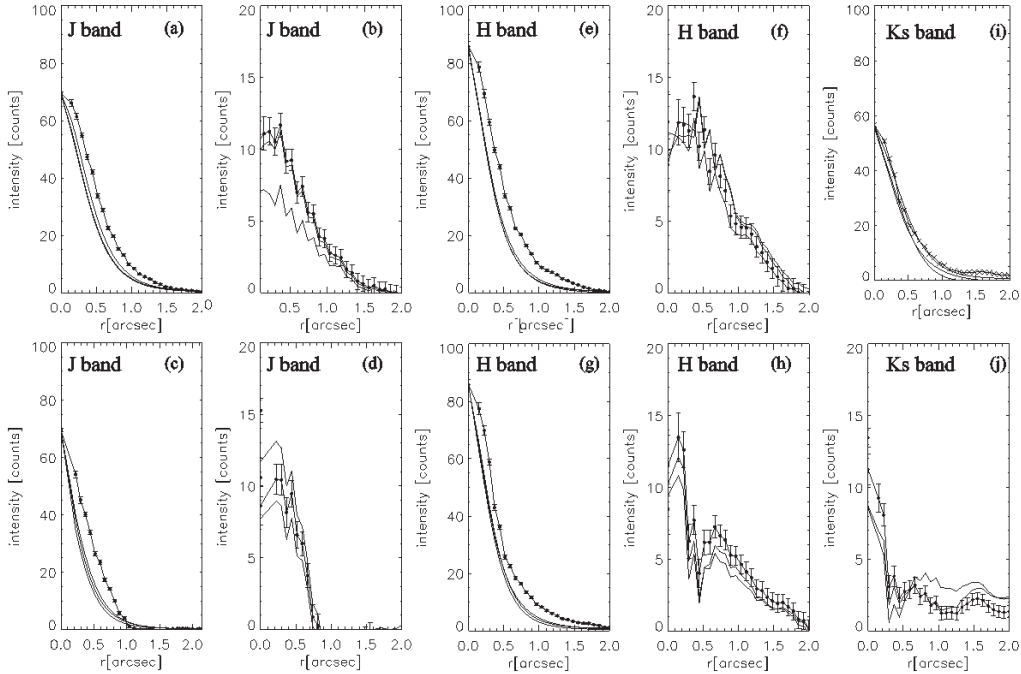}}
\caption{(a) $J$ band radial profile of the BALQ (solid line) and three nearby stars  (dotted, dashed, dash-dot; see Fig. \ref{fig:hbal}). The peak flux of each star is scaled to the peak flux of the BALQ. (b) The residual radial host galaxy profile after subtracting the stellar profile, scaled such that the residual emission is approximately flat close to the center. (c) and (d) correspond to (a) and (b) but for a cut along a position angle of 132$^\circ$. (e, f) and (g, h) same as (a, b) and (c, d) but for $H$ band. (i) is the $Ks$ band radial profile for the position angle of 132$^\circ$ and (j) is the residual after subtraction of the stellar profile. The profile of the BALQ along the PA of 45$^\circ$ shows only a marginal extent with respect to the stellar profile. See text for details.}
\label{fig:radProfile}
\end{center}
\end{figure*}

\section{Observation and Data Reduction}

The data were acquired with the Infrared Spectrometer
and Array Camera (ISAAC, \citealt{1995SPIE.2475..262M}) mounted at the
Nasmyth B focus of Unit Telescope 1 (UT1, Antu) of the Very Large Telescope
(VLT; ESO, Chile). 

Imaging was performed in the broad bands $J$, $H$ and $Ks$ (1.2 - 2.2~$\mu$m).
The 1024x1024 pixel Hawaii Rockwell array detector provides a pixel
scale of 0.1484\arcsec/pixel with a field of view of 152x152~
arcsec$^2$. On source integration time was 2880s, 2160s, and 1080s in $J$, $H$, and $Ks$ respectively.
The data were reduced with IRAF using standard procedures. In all bands the images seem to be adequately flat with a variation smaller than 3\% over the area of interest. Additionally, the science object was moved across the array, so that these variations are reduced due to the averaging of the individual exposures. Successive object and sky observations were subtracted and shifted to produce sky subtracted and averaged (median) images. Fig. \ref{fig:hbal} shows the $H$-band image including the quasar 3C~48. The average seeing was about FWHM$\approx 0\farcs 4$ in $H$ and about 0\farcs 5 in $J$ and $Ks$. The calibration of our data relies on the observation of standard stars \citep{2001MNRAS.325..563H,1998AJ....116.2475P}. More details on the data reduction can be found in \cite{2004A&A...414..919Z}.

\begin{table*}[ht!]
\caption{Photometry of the BALQ and the extended emission. The first part lists $J$, $H$, and $Ks$ magnitudes measured from the radial profiles and in a 1\farcs 5 diameter aperture. The second part lists the deduced colors. Total fluxes and colors are listed in columns (2) and (3). Columns (4) and (5) give the host magnitudes and colors, and (6) and (7) correspondingly for the subtracted nucleus.}
\label{tab:balPhoto}
\begin{center}
\begin{tabular}{ccccccc}
\hline
\hline
Filter      & \multicolumn{2}{c}{Total} & \multicolumn{2}{c}{Host} & \multicolumn{2}{c}{Nucleus}\\
           & \multicolumn{2}{c}{[mag]} & \multicolumn{2}{c}{[mag]} & \multicolumn{2}{c}{[mag]}\\
   & rad. prof. & 1\farcs 5 ap.& rad. prof. & 1\farcs 5 ap. & rad. prof. & 1\farcs 5 ap.\\
(1) & (2) & (3) & (4) & (5) & (6) & (7)\\
\hline
$J$   & $19.9\pm 0.1$ &$20.2\pm 0.1$ & $23.5\pm 0.2$ & $22.6\pm 0.1$ & $22.1\pm 0.2$ & $20.3\pm 0.1$\\
$H$   & $19.1\pm 0.1$ &$19.4\pm 0.1$ & $22.4\pm 0.1$ & $21.4\pm 0.1$ & $21.2\pm 0.1$ & $19.5\pm 0.1$\\
$Ks$  & $18.5\pm 0.1$ &$18.9\pm 0.1$ & $21.8\pm 0.2$ & $20.7\pm 0.1$ & $20.6\pm 0.2$ & $19.1\pm 0.1$\\
\hline
\multicolumn{7}{c}{Colors}\\
\hline
$J-H$   & $0.8\pm 0.1$ & $0.8\pm 0.1$ & $1.1\pm 0.2$ & $1.2\pm 0.1$  & $0.9\pm 0.2$ & $0.8\pm 0.1$ \\
$H-Ks$  & $0.6\pm 0.1$ & $0.5\pm 0.1$ & $0.6\pm 0.2$ & $0.7\pm 0.1$  & $0.6\pm 0.2$ & $0.4\pm 0.1$ \\
$B-Ks$  & 3.2          &  -- &       --     & --   & --           & --\\
\hline
\end{tabular}
\end{center}
\end{table*}

\section{Results and discussion}
While the connection between the super massive black holes (SMBH) and their host galaxies is still not clearly understood, the host galaxies are extremely difficult to observe at redshifts $z>1$, because of the high contrast between nucleus and host galaxy and the surface brightness dimming. The high sensitivity of the VLT and its instrumentation as well as the good seeing conditions (FWHM $<0\farcs 5$) enabled us to investigate the extended rest-frame optical emission around the BALQ~0134+3252, providing a view on the unobscured star formation in these galaxies.

We measure fluxes in $J$, $H$, and $Ks$ of the BALQ, the potential host galaxy and the nucleus. We do not correct for Galactic extinction, because its values are smaller than the photometric error.
Our results are listed Table \ref{tab:balPhoto}. Previous measurements by \cite{1994PASP..106..646K} and \citetalias{1998AJ....115..890C} are given in Table \ref{tab:prevBalPhoto}.

\begin{figure*}[ht!]
\resizebox{\hsize}{!}{\includegraphics{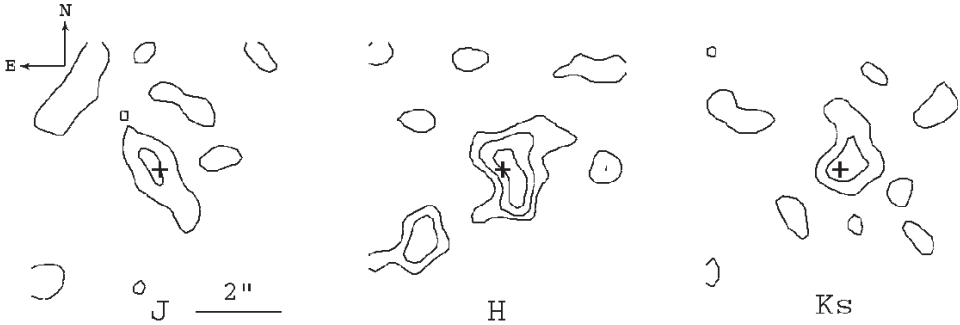}}
\caption{PSF-subtracted $J$, $H$, and $Ks$ images. After subtraction the images have been smoothed with a three pixel Gaussian. The contour levels correspond to $(1,\ 4)\times\sigma$ in $J$, $(1.5,\ 3,\ 4)\times\sigma$ in $H$, and $(2,\ 3)\times\sigma$ in $Ks$ of the sky noise. The crosses shows the position of the nucleus from which the PSF is subtracted.}
\label{fig:residual}
\end{figure*}

\subsection{Overall properties}
Measuring the fluxes in a 2\farcs 6 diameter aperture centered on the quasar nucleus (as done by \citetalias{1998AJ....115..890C}), we find the overall colors of the BALQ to be $J-H=0.8\pm 0.1$ and $H-K=0.5\pm 0.1$ (cf. Fig. \ref{fig:2color}). Using the F555W and the $g$ magnitude together with the transformation from $g$ to $B$ magnitudes ($g=V-0.19+0.41[B-V]$) stated in \cite{1994PASP..106..646K}, we get an observed $B-K\approx 3.3$. This is only slightly reddened compared to a typical $B-K=3$ for unreddened $z>1$ quasars \citep[][~and references therein]{1997ApJ...484L..17H}. \citetalias{1998AJ....115..890C} found a small reddening of the rest-frame UV spectrum of the BALQ compared to the HST composite quasar spectrum \citep{1997ApJ...475..469Z}. Estimating this reddening to be about a factor of 1.3 and following the UV-optical extinction law \citep[e.g.~][]{2005ApJ...625..167S}, the small UV-reddening contributes only little to the rest-frame optical, resulting in the slightly reddened observed $B-K$ color.

In the untreated $H$-band image (Fig. \ref{fig:hbal}) a non stellar-like shape of the BALQ can already be emphasized, which we will discuss next.

\subsection{Radial profiles and 2-dimensional images}
In order to get an improved impression of the faint host galaxy emission we extracted two radial profiles orthogonal to each other in each band, with one along the primary extended component at a position angle (PA) of about $45^\circ$. The profiles are measured over cone-like sections with a cone angle of 60$^\circ$ along this extension (apparent major axis) and along  the orthogonal axis at PA$= 132^\circ$ (cf. Fig. \ref{fig:hbal}).

The radial profiles of the BALQ are compared with the profiles of three nearby  unsaturated stars, circular in appearance (\#1, \#2, and \#3 in Fig. \ref{fig:hbal}), where the stellar profiles are scaled to the QSO peak flux (Figs. \ref{fig:radProfile} (a, c), (e, g), and (i, j)). It is evident from the radial profiles that the intensity distribution in $J$ and $H$ is extended compared to the stellar radial profiles. Next, we fitted a Moffat function, which well represents the ISAAC seeing-limited PSF, to each stellar profile. After subtracting the three functions from the galaxy profile individually, a residual emission is clearly visible in the $J$ and $H$ band (Figs. \ref{fig:radProfile} (b, d) and (f, h) ), whereas in $Ks$ there is only a faint and noisy residual profile (Fig. \ref{fig:radProfile} (j) ) visible. The low signal-to-noise ratio (SNR) in the $Ks$-band residual (especially in the PA$=45^\circ$ cone section (cf. Fig. \ref{fig:residual})) can be accounted for by the facts, that young stellar populations are fading towards longer rest-frame optical wavelengths \citep[e.g. Fig. 8 of][]{2004MNRAS.352..399J} and that the  sky background increases towards the observed $Ks$. In favor of clarity for the reader we only show the error bars of the 'middle' profile (star \#3) which we also used to estimate the flux. The stellar flux was scaled such that its subtraction results in a smooth and flat central part of the profile \citep[cf. discussion in ][]{2001ApJ...546..782M}. This presents a conservative lower limit for the potential emission of a host galaxy. A subtraction of a larger point-like contribution resulting in a decrease of the intensity towards the center is likely to be not physical, and would yield colors not consistent with those of normal galaxies.

\begin{figure}[h!]
\begin{center}
\resizebox{\hsize}{!}{\includegraphics[width=7cm]{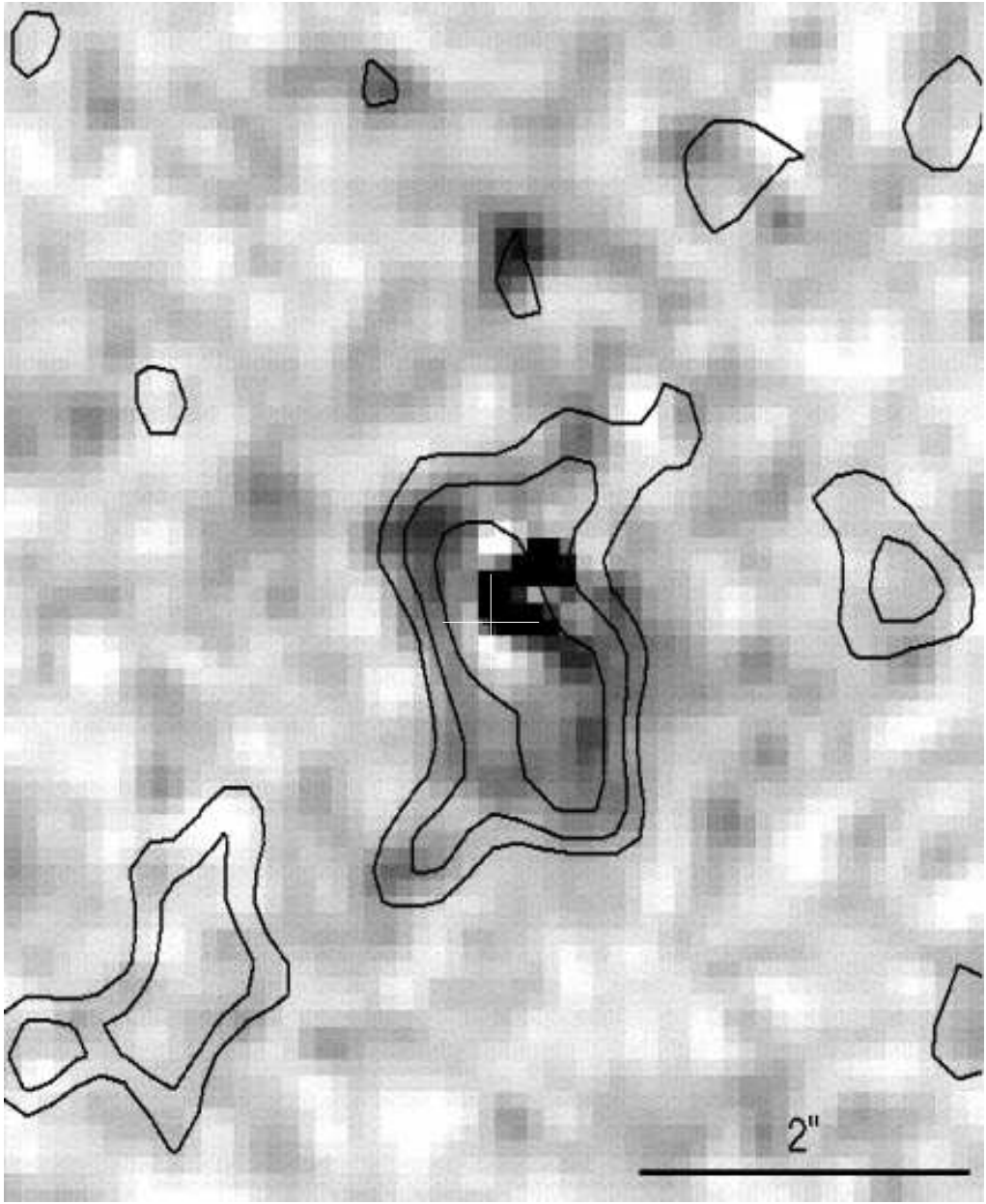}}
\end{center}
\caption{PSF-subtracted HST F814W image from \citetalias{1998AJ....115..890C} overlaid with contours of our three-pixel-smoothed, PSF-subtracted $H$ band image. The lowest contour level corresponds to $1.5\times\sigma$ of the sky noise. The white cross shows the position of the nucleus in our $H$ band image. North is up and east is to the right. Note that the PSF-subtracted HST images is extracted from \citetalias{1998AJ....115..890C}. Despite the values given in Table \ref{tab:balPhoto} no further photometric information is available.}
\label{fig:canalizo_h_bal}
\end{figure}

We measure the fluxes within the two perpendicular cone-sections by summing over the profiles. Then we calculate the mean flux as the geometric mean of the two measurements, because assuming an elliptical morphology we would overestimate the flux when using the major axis alone, and underestimate the flux when only considering the minor axis. The integrated fluxes in our $H$ and $Ks$ measurements are consistent with the limits given by \citetalias{1998AJ....115..890C}. From the residual fluxes we can estimate colors of the extended emission around the BALQ. 

Similarly, we use the same star (\#3) and scaling for the subtraction of the nucleus in the 2-dimensional (2D) images. The results were smoothed with a 3 pixel FWHM Gaussian in order to reduce the noise. Nevertheless, the images show quite strong fluctuation, whereas the $H$ band image has the best signal-to-noise ratio. We measure the fluxes in 1\farcs 5 diameter apertures, because up to this size the fluctuations are not significant. The given errors only include measurement uncertainties. The host contribution to the total light in the NIR is of the order of 10\%. The PSF subtracted images presented in Fig. \ref{fig:residual} show an overall elliptical morphology. The lowest contours correspond to $\sim 2\sigma$ of the sky noise. The structures at distances larger than 3\arcsec from the nucleus are most probably due to noise. 
The feature about 2\arcsec southeast of the nucleus ($H$ and $Ks$) is also visible in the untreated image (inset in Fig. \ref{fig:hbal}) although it is not clear whether it is connected to the BALQ at all. Fig. \ref{fig:canalizo_h_bal} shows our PSF-subtracted $H$ band image in contours overlaid on the PSF-subtracted HST $V$-band image extracted from \citetalias{1998AJ....115..890C}. Although the exact position of the nucleus in the HST image is not known to us, the overall size and orientation is similar to what is reported on the optical by \citetalias{1998AJ....115..890C}.

\begin{figure*}
\begin{center}
\sidecaption
\includegraphics[width=12cm]{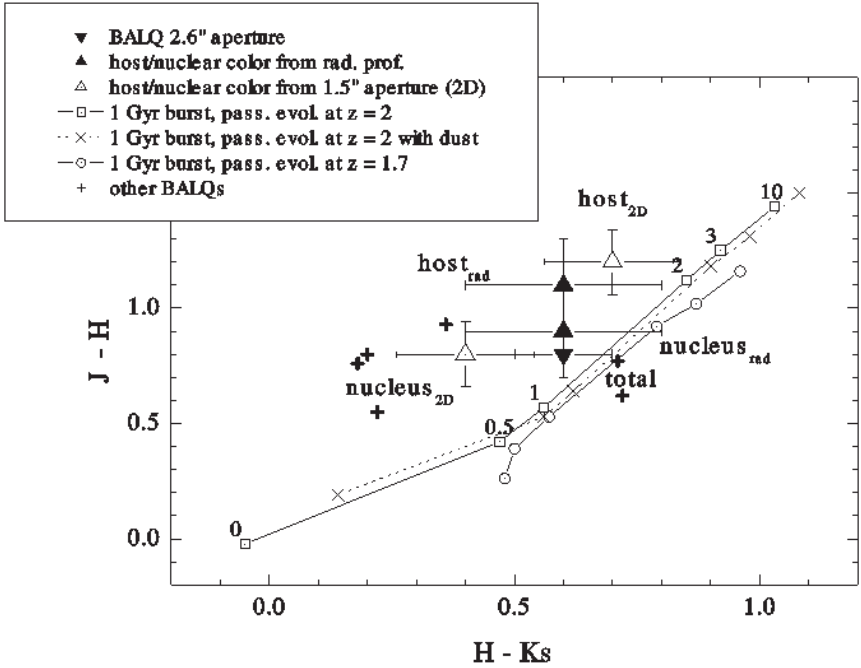}
\caption{Two-color diagram presenting the ($J-H$) vs. ($H-Ks$) observed colors of the BALQ (upside-down triangle), the host, and the nucleus (filled triangles) as discussed in the text. The hollow triangles represent the host and nuclear colors measured in a 1\farcs 5 diameter aperture from the 2D PSF-subtracted images. Also shown are colors of BALQs (crosses) from \cite{1997ApJ...484L..17H}. For comparison colors of a 1~Gyr starburst followed by passive evolution for ages of 0, 0.5, 1, 2, 3, and 10 Gyr at redshifts $z=2$ (open rectangles) and $z=1.7$ (open circles) are displayed \citep[using][]{2003MNRAS.344.1000B}. The dotted line shows the effect of reddening of the $z=2$ track following the model of \cite{2000ApJ...539..718C} with $\tau_V=1.5$ and $\mu=0.3$.}
\label{fig:2color}
\end{center}
\end{figure*}

The colors of the BALQ, its host, and its nucleus together with colors of starbursts of different ages at $z=2$ and $z=1.7$ are presented in Fig. \ref{fig:2color}. For comparison colors of other BALQs listed in \cite{1997ApJ...484L..17H} are displayed. Among them are Hawaii~167 \citep{1994ApJ...432L..83C} and the radio-loud QSO 1556+3517 \citep{1997ApJ...484L..17H}, two of the reddest ($B-K>5$) quasars known. Within the errors the colors of the two methods (radial profiles and 2D photometry) are consistent with each other.

\begin{figure*}
\begin{center}
\sidecaption
\includegraphics[width=12cm]{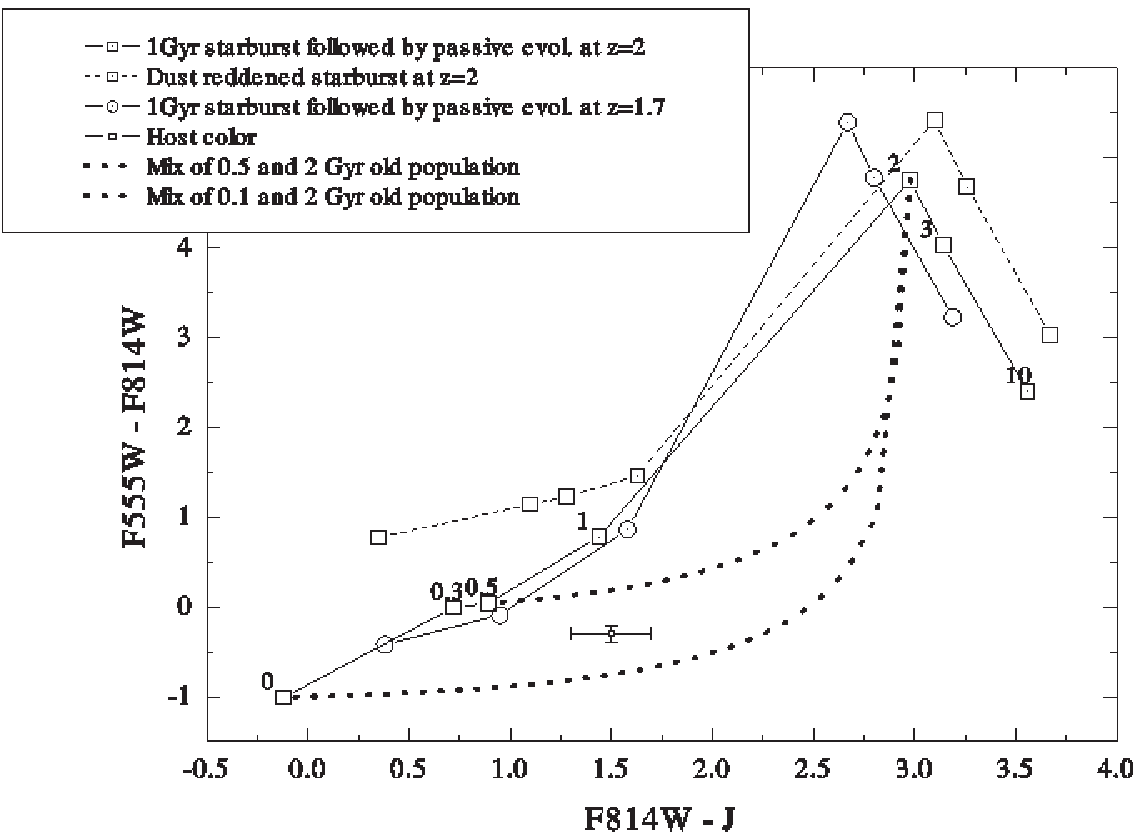}
\caption{Two-color diagram presenting the ($F555W-F814W$) vs. ($F814W-J$) observed colors of the BALQ host galaxy (cf Fig.~\ref{fig:2color}). Also shown are the colors of 1~Gyr single starbursts followed by passive evolution at $z=2$ (open squares; numbers indicate the age in Gyr) and $z=1.7$ (open circles). The effect of reddening of the $z=2$ track, with $\tau_V=1.5$ and $\mu=0.3$ according to the model of \cite{2000ApJ...539..718C}, is shown as open squares joined by a dashed line. Furthermore two curves displaying the effect of mixing of a 0.1~Gyr and 2~Gyr old (lower dotted line) and of a 0.5~Gyr and 2~Gyr old (upper dotted line) population are shown.}
\label{fig:2color2}
\end{center}
\end{figure*}

Combining the available photometric data, we use the four colors ($F555W-F814W$), ($F814W-J$), ($J-H$), and ($H-Ks$) to get a first impression of the stellar content of the extended emission by comparison with colors of simple stellar populations (Figs. \ref{fig:2color} and \ref{fig:2color2}). We calculated colors of a 1~Gyr starburst followed by passive evolution at redshifts of $z=2$ and $z=1.7$ using the population synthesis code of \cite{2003MNRAS.344.1000B}, and utilizing the Padova1994 initial spectral energy distributions of solar metalicity with a Chabrier initial mass function \citep{2003PASP..115..763C}.

The host colors from Fig. \ref{fig:2color} infer a stellar population of about 2~Gyr dominating the emission at the redshifts of the BALQ.
At the redshift of the intermediate absorber, the ages are somewhat older. 
Thus, we see an older stellar population in the rest-frame optical than did \citetalias{1998AJ....115..890C} in the rest-frame UV only, which is typically dominated by young and hot stars. The view on the unobscured star formation can be seen in Fig. \ref{fig:2color2}. The rest-frame UV/optical host colors lie well below the color track of the 1~Gyr starburst. We calculated the mixing curve for a composite color made up of a 0.1 or 0.5~Gyr and a 2~Gyr population (the two thick dotted curves in Fig. \ref{fig:2color2}). It is evident that a young population dominates the rest-frame UV/optical emission, accounting for about 80\% of the colors. Correspondingly, the dominating population would be somewhat younger at $z=1.667$. Similar results are found by \citet[][ their Fig. 6]{2004ApJ...614..568J} in their study of likewise luminous $z\sim 2-3$ GEMS quasar hosts.

Further details on the stellar populations are difficult to obtain, because of the possible degeneracies of colors related to dust and metalicity \citep[cf.][]{2004ApJ...614..568J}. Moreover, the noisy residual after subtraction of the QSO nucleus can influence the resulting colors.

\begin{table}[ht!]
\caption{Physical scales of the residual emission determined from the $H$-band radial profile. 
}
\label{tab:balsize}
\begin{center}
\begin{tabular}{ccccc}
\hline
\hline
HWHM$_\mathrm{major}$ & HWHM$_\mathrm{minor}$ & $r_{1/2}$ & [kpc/\arcsec]  & [kpc/\arcsec] \\
~[\arcsec] & [\arcsec] & [\arcsec] & ($z=2$) & ($z=1.7$)\\
(1) & (2) & (3) & (4) & (5)\\
\hline
$0.9\pm 0.1$ &  $0.5\pm 0.3$ & $0.4\pm 0.1$ & 8.3& 8.5\\
\hline
\end{tabular}
\end{center}
\end{table}

Assuming that the extended emission originates at the redshift of the BALQ, $z=2.169$, 1\arcsec corresponds to a linear scale of 8.3~kpc. In the unified scheme BALQs should be common QSOs at the respective redshift, viewed such that our line of sight intercepts the QSO wind. Thus, we can compare our results with hosts of present day and $z\sim 2-3$ radio-loud and radio-quiet QSOs. For this purpose we use two model independent measures of the host galaxy size (Table \ref{tab:balsize}). First we measure the mean half width at half maximum (HWHM) of the radial profile, because there is some degree of degeneracy in the radial profiles (cf. Fig. \ref{fig:radProfile} (h)). Using the $H$ band profile, we find a HWHM radius of $r=0\farcs 9\pm0\farcs 1$, i.e. a physical scale of $(7.5\pm 0.8)$~kpc, along the major axis. For the minor axis we find $r=0\farcs 5\pm0\farcs 3$, corresponding to $(4.2\pm 2.5)$~kpc along the minor axis. In case of an elliptical galaxy this would give a Hubble type E4. For a spiral galaxy this would imply an inclination of about $(60\pm 10)^\circ$. The axis ratio is also in agreement with the 2D PSF-subtracted $H$ band image (Fig. \ref{fig:canalizo_h_bal}).
As a second measure, the half-light radius derived from the $H$ band radial profile is about $r_{1/2}=0\farcs 4\pm 0\farcs 1$, corresponding to $(3.3\pm 0.8)$~kpc. \cite{2001ApJ...550..122R} find a similar physical scale (2.3~kpc) in their sample of radio quiet QSOs at comparable redshift. These quite compact sizes seem to be more typical for Lyman break galaxies at $z\sim 2-3$ \citep{1996ApJ...470..189G,2000RSPTA.358.2001D} than for present day QSOs hosts \citep{1999MNRAS.308..377M,1997ApJ...479..642B}. This could imply that the compact systems are still in the process of forming. Note that because of the low SNR of our data we could be biased towards more compact emission. Furthermore we use the method described in \cite{2004ApJ...614..568J} to estimate an absolute $B$ band magnitude (Table \ref{tab:absMag}). Assuming an optical power law, $f_\lambda\propto \lambda^\beta$, we can find the index $\beta$ from the NIR colors and then calculate the apparent $B$ magnitude. We find $M_B\approx -22.6\pm 0.2$, which is also comparable to the results of \cite{2001ApJ...550..122R}. The possible host would correspond to a $\sim 4\times L_B^*$ galaxy using the luminosity function of local galaxies from \cite{1992ApJ...390..338L}\footnote{Their $M_B^*=-19.5$ for $h=1$ has been adapted to our cosmology, i.e. $M_B^*\approx-21$.}. This is at the upper end of the range of luminosities ($0.2-4L^*_B$) found by \cite{2001ApJ...550..122R}. 

At the redshift of the absorber, $z=1.667$, 1\arcsec corresponds to 8.5~kpc. Thus, the physical scales would correspond to a HWHM$_\mathrm{major}\approx 7.7$~kpc and HWHM$_\mathrm{minor}\approx 4.3$~kpc. The half-light radius would be $r_{1/2}\approx 3.4$~kpc. The absolute $B$ magnitude of $M_B\approx-21.9$ would correspond to a $2.5\times L_B^*$ host galaxy. Together with the colors this is still consistent with a relatively young stellar population at moderate extinction and with the host sizes at these redshifts. We therefore cannot pinpoint, whether the emission is coming from the BALQ host or from the intermediate absorber. 

\begin{table}[h!]
\caption{Estimated absolute $B$-band magnitude from 1\farcs 5 aperture measurement. Given are the values for the redshift of the BALQ ($z=2.169$) and the redshift of the intermediate absorber ($z=1.667$). See text for details.}
\label{tab:absMag}
\begin{center}
\begin{tabular}{cccc}
\hline
\hline
\multicolumn{2}{c}{$M_B$ Host} & \multicolumn{2}{c}{$M_B$ Nucleus} \\
($z=2.169$)   & ($z=1.667$)   & ($z=2.169$)   & ($z=1.667$)\\
(1) & (2) & (3) & (4)\\
\hline
$-22.5\pm 0.2$& $-21.9\pm 0.2$& $-24.9\pm 0.2$& $-24.2\pm 0.2$ \\
\hline
\end{tabular}
\end{center}
\end{table}

\section{Summary}
We presented rest-frame optical extended emission around the BALQ 0134+3252, complementing previous results on the rest-frame UV extended emission \citepalias{1998AJ....115..890C}. The UV/optical colors indicate a mixture of a young $\la 500$~Myr and a further evolved $\sim2$~Gyr old stellar population. As the broadband data are affected by various degeneracies related to dust and metalicities, a reliable age dating is still difficult to accomplish. This is even complicated due to the QSO subtraction in our noisy data. The rest-frame optical morphology is similar to the UV morphology. If associated with the BALQ, the host galaxy would correspond to a present day 4$L_B^*$ host. Its compact size, $r_{1/2}=3.2$~kpc, is comparable to radio quiet QSOs and Lyman break galaxies at $z\sim2-3$, but smaller than present day QSO hosts. The small size, the influence of a young stellar population, and the high luminosity fit well into the framework of hierarchical galaxy formation.
At the redshift of the potential Ly$\alpha$ absorber, the sizes are  somewhat larger and the luminosity would correspond to a present day 2.5$L_B^*$ host. Also the rest-frame optical stellar populations would be evolved slightly more than at the BALQ redshift. The data are still comparable to results found for RQQs and LBGs. We cannot, however, distinguish between the BALQ or Ly$\alpha$ absorber origin of the extended continuum emission.

A more detailed stellar population analysis awaits higher SNR and further spectroscopic observations. These can also help uncovering the yet unclear source of the UV/optical emission.

\begin{acknowledgements}
We would like to thank Julia Scharw\"achter, Rainer Sch\"odel, Sebastian Fischer and Thomas Bertram for fruitful discussions. The comments and suggestions of the anonymous referee were very helpful in revising this paper. This work was supported in part by the Deutsche Forschungsgemeinschaft (DFG) via grant SFB 494.
\end{acknowledgements}

\bibliographystyle{aa}
\bibliography{3018}

\begin{thebibliography}{42}
\expandafter\ifx\csname natexlab\endcsname\relax\def\natexlab#1{#1}\fi

\bibitem[{{Antonucci}(1993)}]{1993ARA&A..31..473A}
{Antonucci}, R. 1993, \araa, 31, 473

\bibitem[{{Archibald} {et~al.}(2002){Archibald}, {Dunlop}, {Jimenez}, {Fria{\c
  c}a}, {McLure}, \& {Hughes}}]{2002MNRAS.336..353A}
{Archibald}, E.~N., {Dunlop}, J.~S., {Jimenez}, R., {et~al.} 2002, \mnras, 336,
  353

\bibitem[{{Bahcall} {et~al.}(1997){Bahcall}, {Kirhakos}, {Saxe}, \&
  {Schneider}}]{1997ApJ...479..642B}
{Bahcall}, J.~N., {Kirhakos}, S., {Saxe}, D.~H., \& {Schneider}, D.~P. 1997,
  \apj, 479, 642

\bibitem[{{Becker} {et~al.}(2000){Becker}, {White}, {Gregg}, {Brotherton},
  {Laurent-Muehleisen}, \& {Arav}}]{2000ApJ...538...72B}
{Becker}, R.~H., {White}, R.~L., {Gregg}, M.~D., {et~al.} 2000, \apj, 538, 72

\bibitem[{{Briggs} {et~al.}(1984){Briggs}, {Turnshek}, \&
  {Wolfe}}]{1984ApJ...287..549B}
{Briggs}, F.~H., {Turnshek}, D.~A., \& {Wolfe}, A.~M. 1984, \apj, 287, 549

\bibitem[{{Bruzual} \& {Charlot}(2003)}]{2003MNRAS.344.1000B}
{Bruzual}, G. \& {Charlot}, S. 2003, \mnras, 344, 1000

\bibitem[{{Canalizo} {et~al.}(1998){Canalizo}, {Stockton}, \&
  {Roth}}]{1998AJ....115..890C}
{Canalizo}, G., {Stockton}, A., \& {Roth}, K.~C. 1998, \aj, 115, 890 (CS98)

\bibitem[{{Chabrier}(2003)}]{2003PASP..115..763C}
{Chabrier}, G. 2003, \pasp, 115, 763

\bibitem[{{Charlot} \& {Fall}(2000)}]{2000ApJ...539..718C}
{Charlot}, S. \& {Fall}, S.~M. 2000, \apj, 539, 718

\bibitem[{{Cowie} {et~al.}(1994){Cowie}, {Songaila}, {Hu}, {Egami}, {Huang},
  {Pickles}, {Ridgway}, {Wainscoat}, \& {Weymann}}]{1994ApJ...432L..83C}
{Cowie}, L.~L., {Songaila}, A., {Hu}, E.~M., {et~al.} 1994, \apjl, 432, L83

\bibitem[{{Dickinson}(2000)}]{2000RSPTA.358.2001D}
{Dickinson}, M. 2000, in Astronomy, physics and chemistry of $H^{+}_{3}$, 2001

\bibitem[{{Egami} {et~al.}(1996){Egami}, {Iwamuro}, {Maihara}, {Oya}, \&
  {Cowie}}]{1996AJ....112...73E}
{Egami}, E., {Iwamuro}, F., {Maihara}, T., {Oya}, S., \& {Cowie}, L.~L. 1996,
  \aj, 112, 73

\bibitem[{{Gallagher} {et~al.}(2004){Gallagher}, {Brandt}, {Chartas},
  {Garmire}, \& {Sambruna}}]{2004AdSpR..34.2594G}
{Gallagher}, S.~C., {Brandt}, W.~N., {Chartas}, G., {Garmire}, G.~P., \&
  {Sambruna}, R.~M. 2004, Advances in Space Research, 34, 2594

\bibitem[{{Giavalisco} {et~al.}(1996){Giavalisco}, {Steidel}, \&
  {Macchetto}}]{1996ApJ...470..189G}
{Giavalisco}, M., {Steidel}, C.~C., \& {Macchetto}, F.~D. 1996, \apj, 470, 189

\bibitem[{{Hall} {et~al.}(1997){Hall}, {Martini}, {Depoy}, \&
  {Gatley}}]{1997ApJ...484L..17H}
{Hall}, P.~B., {Martini}, P., {Depoy}, D.~L., \& {Gatley}, I. 1997, \apjl, 484,
  L17

\bibitem[{{Hamann} \& {Sabra}(2004)}]{2004ASPC..311..203H}
{Hamann}, F. \& {Sabra}, B. 2004, in Astronomical Society of the Pacific
  Conference Series, 203

\bibitem[{{Hawarden} {et~al.}(2001){Hawarden}, {Leggett}, {Letawsky},
  {Ballantyne}, \& {Casali}}]{2001MNRAS.325..563H}
{Hawarden}, T.~G., {Leggett}, S.~K., {Letawsky}, M.~B., {Ballantyne}, D.~R., \&
  {Casali}, M.~M. 2001, \mnras, 325, 563

\bibitem[{{Hazard} {et~al.}(1984){Hazard}, {Morton}, {Terlevich}, \&
  {McMahon}}]{1984ApJ...282...33H}
{Hazard}, C., {Morton}, D.~C., {Terlevich}, R., \& {McMahon}, R. 1984, \apj,
  282, 33

\bibitem[{{Hewett} \& {Foltz}(2003)}]{2003AJ....125.1784H}
{Hewett}, P.~C. \& {Foltz}, C.~B. 2003, \aj, 125, 1784

\bibitem[{{Jahnke} {et~al.}(2004{\natexlab{a}}){Jahnke}, {Kuhlbrodt}, \&
  {Wisotzki}}]{2004MNRAS.352..399J}
{Jahnke}, K., {Kuhlbrodt}, B., \& {Wisotzki}, L. 2004{\natexlab{a}}, \mnras,
  352, 399

\bibitem[{{Jahnke} {et~al.}(2004{\natexlab{b}}){Jahnke}, {S{\' a}nchez},
  {Wisotzki}, {Barden}, {Beckwith}, {Bell}, {Borch}, {Caldwell}, {H{\"
  a}ussler}, {Heymans}, {Jogee}, {McIntosh}, {Meisenheimer}, {Peng}, {Rix},
  {Somerville}, \& {Wolf}}]{2004ApJ...614..568J}
{Jahnke}, K., {S{\' a}nchez}, S.~F., {Wisotzki}, L., {et~al.}
  2004{\natexlab{b}}, \apj, 614, 568

\bibitem[{{Kirhakos} {et~al.}(1994){Kirhakos}, {Sargent}, {Schneider},
  {Bahcall}, {Jannuzi}, {Maoz}, \& {Small}}]{1994PASP..106..646K}
{Kirhakos}, S., {Sargent}, W.~L.~W., {Schneider}, D.~P., {et~al.} 1994, \pasp,
  106, 646

\bibitem[{{Lewis} {et~al.}(2003){Lewis}, {Chapman}, \&
  {Kuncic}}]{2003ApJ...596L..35L}
{Lewis}, G.~F., {Chapman}, S.~C., \& {Kuncic}, Z. 2003, \apjl, 596, L35

\bibitem[{{Lipari}(1994)}]{1994ApJ...436..102L}
{Lipari}, S. 1994, \apj, 436, 102

\bibitem[{{Loveday} {et~al.}(1992){Loveday}, {Peterson}, {Efstathiou}, \&
  {Maddox}}]{1992ApJ...390..338L}
{Loveday}, J., {Peterson}, B.~A., {Efstathiou}, G., \& {Maddox}, S.~J. 1992,
  \apj, 390, 338

\bibitem[{{Low} {et~al.}(1989){Low}, {Cutri}, {Kleinmann}, \&
  {Huchra}}]{1989ApJ...340L...1L}
{Low}, F.~J., {Cutri}, R.~M., {Kleinmann}, S.~G., \& {Huchra}, J.~P. 1989,
  \apjl, 340, L1

\bibitem[{{McLeod} \& {McLeod}(2001)}]{2001ApJ...546..782M}
{McLeod}, K.~K. \& {McLeod}, B.~A. 2001, \apj, 546, 782

\bibitem[{{McLure} {et~al.}(1999){McLure}, {Kukula}, {Dunlop}, {Baum}, {O'Dea},
  \& {Hughes}}]{1999MNRAS.308..377M}
{McLure}, R.~J., {Kukula}, M.~J., {Dunlop}, J.~S., {et~al.} 1999, \mnras, 308,
  377

\bibitem[{{Moorwood}(1995)}]{1995SPIE.2475..262M}
{Moorwood}, A.~F. 1995, in Proc. SPIE Vol. 2475, p. 262-267, Infrared Detectors
  and Instrumentation for Astronomy, Albert M. Fowler; Ed., Vol. 2475, 262--267

\bibitem[{{Page} {et~al.}(2001){Page}, {Stevens}, {Mittaz}, \&
  {Carrera}}]{2001Sci...294.2516P}
{Page}, M.~J., {Stevens}, J.~A., {Mittaz}, J.~P.~D., \& {Carrera}, F.~J. 2001,
  Science, 294, 2516

\bibitem[{{Persson} {et~al.}(1998){Persson}, {Murphy}, {Krzeminski}, {Roth}, \&
  {Rieke}}]{1998AJ....116.2475P}
{Persson}, S.~E., {Murphy}, D.~C., {Krzeminski}, W., {Roth}, M., \& {Rieke},
  M.~J. 1998, \aj, 116, 2475

\bibitem[{{Reichard} {et~al.}(2003){Reichard}, {Richards}, {Hall}, {Schneider},
  {Vanden Berk}, {Fan}, {York}, {Knapp}, \& {Brinkmann}}]{2003AJ....126.2594R}
{Reichard}, T.~A., {Richards}, G.~T., {Hall}, P.~B., {et~al.} 2003, \aj, 126,
  2594

\bibitem[{{Richards} {et~al.}(2003){Richards}, {Hall}, {Vanden Berk},
  {Strauss}, {Schneider}, {Weinstein}, {Reichard}, {York}, {Knapp}, {Fan},
  {Ivezi{\' c}}, {Brinkmann}, {Budav{\' a}ri}, {Csabai}, \&
  {Nichol}}]{2003AJ....126.1131R}
{Richards}, G.~T., {Hall}, P.~B., {Vanden Berk}, D.~E., {et~al.} 2003, \aj,
  126, 1131

\bibitem[{{Ridgway} {et~al.}(2001){Ridgway}, {Heckman}, {Calzetti}, \&
  {Lehnert}}]{2001ApJ...550..122R}
{Ridgway}, S.~E., {Heckman}, T.~M., {Calzetti}, D., \& {Lehnert}, M. 2001,
  \apj, 550, 122

\bibitem[{{Sofia} {et~al.}(2005){Sofia}, {Wolff}, {Rachford}, {Gordon},
  {Clayton}, {Cartledge}, {Martin}, {Draine}, {Mathis}, {Snow}, \&
  {Whittet}}]{2005ApJ...625..167S}
{Sofia}, U.~J., {Wolff}, M.~J., {Rachford}, B., {et~al.} 2005, \apj, 625, 167

\bibitem[{{Spergel} {et~al.}(2003){Spergel}, {Verde}, {Peiris}, {Komatsu},
  {Nolta}, {Bennett}, {Halpern}, {Hinshaw}, {Jarosik}, {Kogut}, {Limon},
  {Meyer}, {Page}, {Tucker}, {Weiland}, {Wollack}, \&
  {Wright}}]{2003ApJS..148..175S}
{Spergel}, D.~N., {Verde}, L., {Peiris}, H.~V., {et~al.} 2003, \apjs, 148, 175

\bibitem[{{Stocke} {et~al.}(1992){Stocke}, {Morris}, {Weymann}, \&
  {Foltz}}]{1992ApJ...396..487S}
{Stocke}, J.~T., {Morris}, S.~L., {Weymann}, R.~J., \& {Foltz}, C.~B. 1992,
  \apj, 396, 487

\bibitem[{{Voit} {et~al.}(1993){Voit}, {Weymann}, \&
  {Korista}}]{1993ApJ...413...95V}
{Voit}, G.~M., {Weymann}, R.~J., \& {Korista}, K.~T. 1993, \apj, 413, 95

\bibitem[{{Weymann} {et~al.}(1991){Weymann}, {Morris}, {Foltz}, \&
  {Hewett}}]{1991ApJ...373...23W}
{Weymann}, R.~J., {Morris}, S.~L., {Foltz}, C.~B., \& {Hewett}, P.~C. 1991,
  \apj, 373, 23

\bibitem[{{Willott} {et~al.}(2003){Willott}, {Rawlings}, \&
  {Grimes}}]{2003ApJ...598..909W}
{Willott}, C.~J., {Rawlings}, S., \& {Grimes}, J.~A. 2003, \apj, 598, 909

\bibitem[{{Zheng} {et~al.}(1997){Zheng}, {Kriss}, {Telfer}, {Grimes}, \&
  {Davidsen}}]{1997ApJ...475..469Z}
{Zheng}, W., {Kriss}, G.~A., {Telfer}, R.~C., {Grimes}, J.~P., \& {Davidsen},
  A.~F. 1997, \apj, 475, 469

\bibitem[{{Zuther} {et~al.}(2004){Zuther}, {Eckart}, {Scharw{\" a}chter},
  {Krips}, \& {Straubmeier}}]{2004A&A...414..919Z}
{Zuther}, J., {Eckart}, A., {Scharw{\" a}chter}, J., {Krips}, M., \&
  {Straubmeier}, C. 2004, \aap, 414, 919

\end{thebibliography}
\
\end{document}